\documentclass{article}
\usepackage{graphicx}
\begin{document}

\begin{center}
{\LARGE \bf Wigner's new physics frontier: Physics of two-by-two matrices,
including the Lorentz group and optical instruments} \\

\vspace{7mm}

Sibel Ba{\c s}kal \footnote{electronic
address:baskal@newton.physics.metu.edu.tr} \\
Department of Physics, Middle East Technical University,\\
06531 Ankara, Turkey
\vspace{5mm}

Elena Georgieva\footnote{electronic address:
egeorgie@pop500.gsfc.nasa.gov}\\
Science Systems and Applications, Inc., Lanham, MD 20771, U.S.A., and \\
National Aeronautics and Space Administration, \\Goddard Space
Flight Center, Laser and Electro-Optics Branch,\\ Code 554, Greenbelt,
Maryland 20771, U.S.A.

\vspace{5mm}

Y. S. Kim\footnote{electronic address: yskim@physics.umd.edu}\\
Department of Physics, University of Maryland,\\
College Park, Maryland 20742, U.S.A.\\

\end{center}

\section*{Abstract}

According to Eugene Wigner, quantum mechanics is a physics of Fourier
transformations, and special relativity is a physics of Lorentz
transformations.  Since two-by-two matrices with unit determinant form
the group $SL(2,c)$ which acts as the universal covering group of the
Lorentz group, the two-by-two matrices constitute the natural language
for special relativity.  The central language for optical instruments
is the two-by-two matrix called the beam transfer matrix, or the so-called
$ABCD$ matrix.
It is shown that the $ABCD$ matrices also form the $SL(2,C)$ group.
Thus, it is possible to perform experiments in special relativity using
optical instruments.  Likewise, the optical instruments can be
explained in terms of the symmetry of relativistic particles.

Based on this review article, the last author(YSK) presented papers
at a number of conferences, including the 8th International Wigner
Symposium (New York, U.S.A., 2003), the 8th International Conference
on Squeezed States and Uncertainty Relations (Puebla, Mexico, 2003),
the Symmetry Festival (Budapest, Hungary, 2003), and the International
Conference on Physics and Control (Saint Petersburg, Russia, 2003).

\newpage
\section{Introduction}\label{intro}

Eugene Paul Wigner received the 1963 Noble prize in physics for his
contributions to symmetry problems in physics.  Even before the
formulation of the present form of quantum mechanics, Wigner perceived
the importance of symmetry problems in quantum mechanics.  In his
book which was published in 1931~\cite{wig31}, Wigner formulated
the quantum theory
of angular momentum in terms of the three-dimensional rotational
symmetries.  In so doing, he introduced group theoretical methods
to physics.

In 1932~\cite{wig32}, he published a paper concerning thermodynamic
corrections to equilibrium systems, and introduced a phase-space picture
of quantum mechanics.  The phase-space distribution function he introduced
in this paper is called the Wigner function.  The Wigner function is an
indispensable theoretical tool in quantum optics and in the foundations of
quantum mechanics~\cite{knp91}.

Frederick Seitz was Wigner's first student at Princeton. In their papers
published in 1933 and 1934~\cite{wig33}, they worked out the symmetry
of sodium crystals, in terms of group theory.  They initiated application
of quantum mechanics to matter.  This field is now known as condensed
matter physics.

By 1936, the spin and isospin symmetries for nucleons were well
established in physics.  Wigner was interested whether those two
separate symmetries come from one big symmetry.  In so doing, he formulated
the concept of supermultiplets~\cite{wig37}.  Wigner's supermultiplet
theory was the grandfather of the quark model which Gell-Mann formulated in
1963~\cite{gel63}.

In 1939, Wigner published his most fundamental paper dealing with
space-time symmetries of relativistic particles~\cite{wig39}.  In this
paper, Wigner introduced the Lorentz group to physics.  Furthermore,
by introducing his ``little groups,'' Wigner provided the framework for
studying the internal space-time symmetries of relativistic particles.
Since this paper was so ahead of his time, it was rejected by three different
journals before John von Neumann, then the editor of the Annals of
Melodramatics, decided to publish it in his journal.
The scientific contents of this paper have not yet been fully recognized
by the physics community.  We are writing this report as a continuation
of the work Wigner initiated in this history-making paper.

While particle physicists are still struggling to understand internal
space-time symmetries of elementary particles, Wigner's Lorentz group
is becoming useful to many other branches of physics.  Among them is
optical sciences, both quantum and classical.  In quantum optics, the
coherent and squeezed states are representations of the Lorentz
group~\cite{knp91}.  Recently, the Lorentz group is becoming the
fundamental language for classical ray optics.  It is gratifying to note
that optical components, such as lenses, polarizers, interferometers,
lasers, and multi-layers can all be formulated in terms of the Lorentz
group which Wigner formulated in his 1939 paper.  Classical ray optics
is of course a very old subject, but we cannot do new physics without
measurements using optical instruments.
Indeed, classical ray optics constitutes Wigner's frontier in physics.

The word "group theory" sounds like an abstract mathematics, but it is
gratifying that Wigner's little groups can be formulated in
terms of two-by-two matrices, while classical ray optics is largely a
physics of two-by-two matrices.  The mathematical correspondence is
straight-forward.

In order to see this point clearly, let us start with the
following classic example. The second-order differential equation
\begin{equation}
A {d^{2} q(t) \over dt^{2}} + B {d q(t) \over dt} +
C q(t) = F \cos(\omega t) .
\end{equation}
is applicable to a driven harmonic oscillator
with dissipation.  This can also be used for studying an electronic
circuits consisting of inductance, resistance, capacitance, and an
alternator.   Thus, it is possible to study the oscillator system using
the electronic circuit.  Likewise, an algebra of two-by-two matrices
can serve as the scientific language for two more different branches of
physics.

\begin{table}

\caption{Further contents of Einstein's $E = mc^{2}$.  Massive and
massless particles have different energy-momentum relations.  Einstein's
special relativity gives one relation for both.  Wigner's little group
unifies the internal space-time symmetries for massive and massless
particles.  The quark model and the parton model can also be combined into
one covariant model.}\label{einwig}
\begin{center}
\vspace{3mm}
\begin{tabular}{rccc}
\hline \\[-3.9mm]
\hline
{} & {} & {} & {}\\
{} & Massive, Slow \hspace*{1mm} & COVARIANCE \hspace*{1mm}&
Massless, Fast \\[4mm]\hline
{} & {} & {} & {}\\
Energy- & {}  & Einstein's & {} \\
Momentum & $E = p^{2}/2m_{0}$
 & $ E = \sqrt{p^{2}c^{2} + m_{0}^{2}c^{4}}$ & $E = cp$
\\[4mm]\hline
{} & {} & {} & {}\\
Internal & $S_{3} $ & {}  &  $S_{3}$ \\[-1mm]
space-time & {} & Wigner's  & {} \\ [-1mm]
symmetry & $S_{1}, S_{2}$ & Little Group & Gauge
Transformation \\[4mm]\hline
{} & {} & {} & {}\\
Relativistic & {} & Covariant Model  & {} \\[-1mm]
Extended & Quark Model & of    & Partons \\ [-1mm]
Particles & {} & Hadrons & {} \\[2mm]
\hline
\hline
\end{tabular}

\end{center}
\end{table}

There are many physical systems which can be formulated in terms of
two-by-two matrices.  If we restrict that their determinant be one,
there is a well established mathematical descipline called the group
theory of $SL(2,C)$. This aspect was noted in the study of Lorentz
transformations.  In group theoretical terminology, the group $SL(2,C)$
is the universal covering group of the group of the Lorentz group.
In practical terms, to each two-by-two matrix, there corresponds one
four-by-four matrix which performs a Lorentz transformation in the
four-dimensional Minkowskian space.  Thus, if a physical system can
be explained in terms of two-by-two matrices, it can be explained
with the language of Lorentz transformations.  Conversely, the system
can serve as an analogue computer for Lorentz transformations.

Optical filters, polarizers, and interferometers deal with two
independent optical rays.  They superpose the beams, change the
relative phase shift, and change relative amplitudes.   The basic
language here is called the Jones matrix formalism, consisting of
the two-by-two matrix representation of the $SL(2,C)$
group~\cite{hkn97,hkn99}.  The four-by-four Mueller
matrices are derivable from the two-by-two matrices of $SL(2,C)$.

Para-axial lens optics can also be formulated in terms of two-by-two
matrices, applicable to the two-component vector space consisting of
the distance from the optical axis and the slope with respect to the
axis.  The lens and translation matrices are triangular, but they are
basically representations of the $Sp(2)$ group which is the real
subgroup of the group $SL(2,C)$~\cite{bk01,bk03}.

Laser optics is basically multi-lens lens optics.  However, the problem
here is how to get simple mathematical expression for the system of a
large number of the same lens separately by equal distance.  Here again,
group theory simplifies calculations~\cite{bk02}.

In multi-layer optics, we deal with two optical rays moving in
opposite directions.  The standard language in this case is the
S-matrix formalism~\cite{azzam77}. This is also a two-by-two matrix
formalism.  As in the case of laser cavities, the problem is the
multiplication of a large number of matrix chains~\cite{gk01,gk03}.

It is shown in this report that the two-by-two representation of
the six-parameter Lorentz group is the underlying common scientific
language for all of the instruments mentioned above.
While the abstract group theoretical ideas make two-by-two matrix
calculations more systematic and transparent in optics, optical
instruments can act as analogue computers for Lorentz transformations
in special relativity.  It is gratifying to note that special relativity
and ray optics can be formulated as the physics of two-by-two matrices.

In Sec.~\ref{wlittle}, we discuss the historical significance of
Winger's 1939 paper~\cite{wig39} on the Lorentz group and its application
to the internal space-time symmetries of relativistic particles.
In Sec.~\ref{formul}, we present the basic building blocks for the
two-by-two representation of the Lorentz group in terms of the
matrices commonly seen in ray optics. In Secs.~\ref{polari},
\ref{olens}, \ref{mlens}, \ref{lcav}, and \ref{mlayer}, we discuss
polarization optics, one-lens system, multi-lens system, laser
cavities, and multi-layer optics, respectively.

Since the Lorentz group is relatively new to many who study optics,
we explain how it is possible to represent the group using two-by-two
matrices in Appendix~\ref{lorentz}, we explain how the Lorentz group
can be formulated in terms of four-by-four matrices.  It is shown that
the group can have six independent parameters.
In Appendix~\ref{spinor}, we explain how it is possible to formulate
the Lorentz group in terms of two-by-two matrices.  It is shown that
the four-by-four transformation matrices can be constructed from
those two-by-two matrices.
In Appendix~\ref{conju}, it is noted that the four-by-four matrices are
real, their two-by-two counterparts are complex.  However, there is
a three-parameter subgroup called $Sp(2)$.  It is shown that the
complex subgroup $SU(1,1)$ is equivalent to $Sp(2)$ through conjugate
transformation.

The purpose of these appendixes is to give an introduction to group
theoretical methods used in this report and in the recent optics
literature.  In Appendix~\ref{euler}, it is shown that the Lie
group method, in terms of the generators, is not the only method in
constructing group representations.  For  the rotation group and the
three-parameter subgroups of the Lorentz group, it is simpler to
start with the minimum number of starter matrices.  For instance, while there are three generators for the
rotation group in the Lie approach, we can construct the most general
form of the rotation matrix from rotations around two directions,
as Goldstein constructed the Euler angles~\cite{gold80}.

As for the references, we have not made attempts to list all the papers
relevant to the present report.  However the references are given
in the papers in the refereed journals.

\section{Wigner's Little Groups}\label{wlittle}

If the momentum of a particle is much smaller than its rest-mass energy,
the
energy-momentum relation is $E = p^{2}/2m_{0} + m_{0}c^{2}$.
If the momentum
is much larger than its rest-mass energy, the relation is $E = cp$.  These two
different relations can be combined into one covariant formula
$E^{2} = m_{0}^{2}c^{4} + p^{2}c^{2}$.
This aspect of Einstein's $E = mc^{2}$ is also well known.

In addition, particles have internal space-time variables.  Massive
particles have spins while massless particles have their helicities
and gauge variables.  Our first question is whether this aspect of
space-time variables can be unified into one covariant concept.
The answer to this question is Yes.  Wigner's little group does the
job.  In addition, particles can have space-time extensions.  For
instance, in the quark model, hadrons are bound states of quarks.
However, the hadron appears
as a collection of partons when they it moves with speed close to the
velocity of light.  Quarks and partons seem to have quite distinct
properties.  Are they different manifestation of a single covariant
entity?  This is one of the main issues in high-enrgy particles
physics.

By ``further contents'' of Einstein's $E = mc^{2}$, we mean that the
internal space-time structures of massive and massless particles can
be unified into one covariant package, as
$E^{2} = m_{0}^{2}c^{4} + p^{2}c^{2} $
does for the energy-momentum relation.  The mathematical framework of
this program was developed by Eugene Wigner in 1939~\cite{wig39}.
He constructed maximal subgroups of the Lorentz group whose
transformations will leave the four-momentum of a given particle
invariant.  These groups are known as Wigner's little groups.
Thus, the transformations of the little groups change the internal
space-time variables the particle.  The little group is a covariant
entity and takes different forms for the particles moving with
different speeds.

The space-time symmetry of relativistic particles is dictated by
the Poincar\'e group~\cite{wig39}.  The Poincar\'e group is the group
of inhomogeneous Lorentz transformations, namely Lorentz transformations
preceded or followed by space-time translations.
Thus, the Poincar\'e group is a semi-direct product of
the Lorentz and translation groups.  The two Casimir operators of
this group correspond to the (mass)$^{2}$ and (spin)$^{2}$ of a given
particle.  Indeed, the particle mass and its spin magnitude are
Lorentz-invariant quantities.

The question then is how to construct the representations of the
Lorentz group which are relevant to physics.  For this purpose,
Wigner in 1939 studied the maximal subgroups of the Lorentz group
whose transformations leave the four-momentum of a given free
particle~\cite{wig39}.  These subgroups are called the little groups.
Since the little group leaves the four-momentum invariant, it governs
the internal space-time symmetries of relativistic particles.  Wigner
shows in his paper that the internal space-time symmetries of massive
and massless particles are dictated by the little groups which are
locally isomorphic to the three-dimensional rotation group and the
two-dimensional Euclidean groups respectively.

The group of Lorentz transformations consists of three boosts and
three rotations.  The rotations therefore constitute a subgroup of
the Lorentz group.  If a massive particle is at rest, its four-momentum
is invariant under rotations.  Thus the little group for a massive
particle at rest is the three-dimensional rotation group.  Then what is
affected by the rotation?  The answer to this question is very simple.
The particle in general has its spin.  The spin orientation is going
to be affected by the rotation!  If we use the four-vector coordinate
$(t, z, x, y)$, the four-momentum vector for the particle at rest is
$(m_{0}c^{2} ,0 , 0, 0)$, and the three-dimensional rotation group leaves this
four-momentum invariant.  This little group is generated by
\begin{equation}\label{j3}
J_{1} = \pmatrix{0&0&0&0\cr0&0&0&i\cr0&0&0&0\cr0&-i&0&0} , \quad
J_{2} = \pmatrix{0&0&0&0\cr0&0&-i&0\cr0&i&0&0\cr0&0&0&0} , \quad
J_{3} = \pmatrix{0 & 0 & 0 & 0 \cr 0 & 0 & 0 & 0
\cr 0 & 0 & 0 & -i \cr 0 & 0 & i & 0} .
\end{equation}
These are essentially the generators of the three-dimensional rotation
group. They satisfy the commutation relations:
\begin{equation}\label{o3com}
[J_{i}, J_{j}] = i\epsilon_{ijk} J_{k} .
\end{equation}

If the rest-particle is boosted along the $z$ direction, it will pick
up a non-zero momentum component along the same direction.  The above
generators will also be boosted.  The boost will take the form of
conjugation by the boost matrix
\begin{equation}\label{boost}
B = \pmatrix{\cosh\eta &\sinh\eta & 0 & 0\cr
\sinh\eta & \cosh\eta & 0 & 0
\cr 0 & 0 & 1 & 0 \cr 0 & 0 & 0 & 1} .
\end{equation}
This boost will not change the commutation relations of Eq.(\ref{o3com})
for $O(3)$, and the boosted little group will still leave the
boosted four-momentum invariant.  Thus, the little group of a moving
massive particle is still  $O(3)$-like.

It is not possible to bring a massless particle to its rest frame.
In his 1939 paper~\cite{wig39}, Wigner observed that the little group
for a massless particle moving along the $z$ axis is generated by the
rotation generator around the $z$ axis, namely $J_{3}$ of Eq.(\ref{j3}),
and two other generators which take the form
\begin{equation}\label{n1n2}
N_{1} = \pmatrix{0 & 0 & i & 0 \cr 0 & 0 & i & 0
\cr i & -i & 0 & 0 \cr 0 & 0 & 0 & 0} ,  \qquad
N_{2} = \pmatrix{0 & 0 & 0 & i \cr 0 & 0 & 0 & i
\cr 0 & 0 & 0 & 0 \cr i & -i & 0 & 0} .
\end{equation}
If we use $K_{i}$ for the boost generator along the i-th axis, these
matrices can be written as
\begin{equation}
N_{1} = K_{1} - J_{2} , \qquad N_{2} = K_{2} + J_{1} ,
\end{equation}
with
\begin{equation}
K_{1} = \pmatrix{0&0&i&0\cr0&0&0&0\cr i&0&0&0\cr 0&0&0&0} , \qquad
K_{2} = \pmatrix{0&0&0&i\cr0&0&0&0\cr0&0&0&0\cr i&0&0&0} .
\end{equation}
The generators $J_{3}, N_{1}$ and $N_{2}$ satisfy the following set
of commutation relations.
\begin{equation}\label{e2lcom}
[N_{1}, N_{2}] = 0 , \qquad [J_{3}, N_{1}] = iN_{2} ,
\qquad [J_{3}, N_{2}] = -iN_{1} .
\end{equation}

In order to understand the mathematical basis of the above commutation
relations, let us consider transformations on a two-dimensional plane
with the $xy$ coordinate system.  We can then make rotations around
the origin and translations along the $x$ and $y$ directions.  If we
write these generators as $L, P_{x}$ and $P_{y}$ respectively, they
satisfy the commutation relations~\cite{knp86}
\begin{equation}\label{e2com}
[P_{x}, P_{y}] = 0 , \qquad [L, P_{x}] = iP_{y} ,
\qquad [L, P_{y}] = -iP_{x} .
\end{equation}
This is a closed set of commutation relations for the generators of the
$E(2)$ group.  If we replace $N_{1}$ and $N_{2}$ of Eq.(\ref{e2lcom})
by $P_{x}$ and $P_{y}$, and $J_{3}$ by $L$, the commutations relations
for the generators of the $E(2)$-like little group becomes those for
the $E(2)$-like little group.  This is precisely why we say that
the little group for massless particles are like $E(2)$.

It is not difficult to associate the rotation generator $J_{3}$ with
the helicity degree of freedom of the massless particle.   Then what
physical variable is associated with the $N_{1}$ and $N_{2}$
generators?  Indeed, Wigner was the one who discovered the existence
of these generators, but did not give any physical interpretation to
these translation-like generators in his original paper~\cite{wig39}.
For this reason, for many years,
only those representations with the zero-eigenvalues of the $N$
operators were thought to be physically meaningful
representations~\cite{wein64}.  It was not until 1971 when Janner
and Janssen reported that the transformations generated by these
operators are gauge transformations~\cite{janner71,kim97poz}.  The
role of this translation-like transformation has also been studied
for spin-1/2 particles, and it was concluded that the polarization
of neutrinos is due to gauge invariance~\cite{hks82,kim97min}.

The $O(3)$-like little group remains $O(3)$-like when the particle is
Lorentz-boosted.  Then, what happens when the particle speed becomes
the speed of light?  The energy-momentum relation
$E^{2} = m_{0}^{2}c^{4} + p^{2}c^{2}$ become $E = pc$.
Is there then a limiting case of the $O(3)$-like little group?
Since those little groups are
like the three-dimensional rotation group and the two-dimensional
Euclidean group respectively, we are first interested in whether
$E(2)$ can be obtained from $O(3)$.  This will then give a clue to
obtain the $E(2)$-like little group as a limiting case of
$O(3)$-like little group.  With this point in mind, let us look into
this geometrical problem.

In 1953, Inonu and Wigner formulated this problem as the contraction
of $O(3)$ to $E(2)$~\cite{inonu53}.  Let us see what they did.  We
always associate the three-dimensional rotation group with a spherical
surface.  Let us consider a circular area of radius 1 kilometer centered
on the north pole of the earth.  Since the radius of the earth is more
than 6,450 times longer, the circular region appears flat.  Thus, within
this region, we use the $E(2)$ symmetry group.  The
validity of this approximation depends on the ratio of the two radii.

How about then the little groups which are isomorphic to $O(3)$ and
$E(2)$?  It is reasonable to expect that the $E(2)$-like little group
can be obtained as a limiting case for of the $O(3)$-like little group
for massless particles.  In 1981, it was observed by Bacry and
Chang~\cite{bacry68} and by Ferrara and Savoy~\cite{ferrara82}
that this limiting process is the Lorentz boost to infinite-momentum
frame.

In 1983, it was noted by Han {\it et al} that the large-radius limit
in the the contraction of $O(3)$ to $E(2)$ corresponds to the
infinite-momentum limit for the case of $O(3)$-like little group to
$E(2)$-like little group.  They showed that transverse rotation
generators become the generators of gauge transformations in the
limit of infinite momentum~\cite{hks83pl}.

Let us see how this happens. The $J_{3}$ operator of Eq.(\ref{j3}),
which generates rotations around the $z$ axis, is not affected by
the boost conjugation of the $B$ matrix of Eq.(\ref{boost}).  On
the other hand, the $J_{1}$ and $J_{2}$ matrices become
\begin{equation}
N_{1} = \lim_{\eta\rightarrow \infty}e^{-\eta} B^{-1} J_{2} B ,
\qquad N_{2} = \lim_{\eta\rightarrow \infty}-e^{-\eta} B^{-1} J_{1} B ,
\end{equation}
and they become $N_{1}$ and $N_{2}$ given in Eq.(\ref{n1n2}).  The
generators $N_{1}$ and $N_{2}$ are the contracted $J_{2}$ and $J_{1}$
respectively in the infinite-momentum.  In 1987, Kim
and Wigner studied this problem in more detail and showed that the
little group for massless particles is the cylindrical group which is
isomorphic to the $E(2)$ group~\cite{kiwi87jm}.

This completes the second row in Table~\ref{einwig}, where Wigner's
little group
unifies the internal space-time symmetries of massive and massless
particles.  The transverse components of the rotation generators become
generators of gauge transformations in the infinite-momentum limit.

Let us go back to Table I given in Sec.~\ref{intro}.  As for the third
row for relativistic extended particles, the most efficient approach
is to construct representations of the little groups
using the wave functions which can be Lorentz-boosted.  This means
that we have to construct wave functions which are consistent with
all known rules of quantum mechanics and special relativity.  It is
possible to construct
harmonic oscillator wave functions which satisfy these conditions.
We can then take the low-speed and high-speed limits of the covariant
harmonic oscillator wave functions for the quark model and the
parton model  respectively.  This aspect was extensively discussed in
the literature~\cite{knp86}, and is beyond the scope of the present
report.

\section{Formulation of the Problem}\label{formul}
Let us consider two optical beams propagating along the $z$ axis.
We are then led to the column vector:
\begin{equation}\label{jones}
\pmatrix{A_{1}~\exp{\left(-i\left(kz -\omega t + \phi_{1}\right)\right) }
\cr A_{2}~\exp{\left(-i\left(kz -\omega t + \phi_{2}\right)\right) } } .
\end{equation}
We can then achieve a phase shift between the beams by applying the
two-by-two matrix:
\begin{equation}\label{rot11}
\pmatrix{e^{i\phi/2} & 0 \cr 0 & e^{-i\phi/2}}.
\end{equation}
If we are interested in mixing up the two beams, we can apply
\begin{equation}\label{rot22}
\pmatrix{\cos(\theta/2)  & -\sin(\theta/2)  \cr
\sin(\theta/2)  & \cos(\theta/2)}
\end{equation}
to the column vector.

If the amplitudes become changed by either by attenuation or
reflection, we can use the matrix
\begin{equation}\label{boost11}
\pmatrix{e^{\eta/2}  & 0  \cr 0 & e^{-\eta/2}}
\end{equation}
for the change.  In this paper, we are dealing only with the
relative amplitudes, or the ratio of the amplitudes.

Repeated applications of these matrices lead to the form
\begin{equation}\label{alpha}
\pmatrix{\alpha & \beta \cr \gamma  & \delta} ,
\end{equation}
where the elements are in general complex numbers.  The
determinant of this matrix is one.  Thus, the matrix can
have six independent parameters.

Indeed, this matrix is the most general form of the matrices
in the $SL(2,c)$ group, which is known to be the universal
covering group for the six-parameter Lorentz group.  This means
that, to each two-by-two matrix of $SL(2,c)$, there corresponds
one four-by-four matrix of the group of Lorentz transformations
applicable to the four-dimensional Minkowski space~\cite{knp86}.
It is possible to construct explicitly the four-by-four
Lorentz transformation matrix from the parameters $\alpha, \beta,
\gamma,$ and $\delta$.  This expression is available in the
literature~\cite{knp86}, and we consider here only special cases.

As is shown in Appendix~\ref{lorentz}, the Lorentz group includes
Lorentz boosts along three different directions and rotations around
those three different directions.  A rotation around the $z$ axis in the
Minowskian space can be
written as
\begin{equation}
\pmatrix{1 & 0 & 0 & 0 \cr 0 & 1 & 0 & 0 \cr
0 & 0 & \cos\phi & -\sin\phi \cr
0 & 0 & \sin\phi & \cos\phi } .
\end{equation}
This four-by-four matrix corresponds to the two-by-two matrix of
Eq.(\ref{rot22}).  We use in this paper the four-vector convention
$(t, z, x, y)$.

The rotation around the $y$ axis can be written as
\begin{equation}
\pmatrix{1 & 0 & 0 & 0 \cr
0 &  \cos\theta & -\sin\theta & 0 \cr
0 &  \sin\theta & \cos\theta  & 0 \cr
0 & 0 & 0 & 1 } .
\end{equation}
The two-by-two matrix of Eq.(\ref{boost11}) corresponds to the Lorentz
boost matrix
\begin{equation}
\pmatrix{\cosh\eta & \sinh\eta & 0 & 0 \cr
\sinh\eta & \cosh\eta & 0 & 0 \cr
0 &  0 & 1 & 0 \cr 0 & 0 & 0 & 1 } .
\end{equation}

The four-by-four representation of the Lorentz group can be constructed
from the two-by-two representation of the $SL(2,c)$ group, which is known
as the universal covering group of the six-parameter Lorentz group.  This
aspect is discussed in Appendix~\ref{spinor}.

Let us go back to Eq.(\ref{alpha}), the $SL(2,c)$ group represented by
this matrix has many interesting subgroups.  If
the matrices are to be
Hermitian, then the subgroup is $SU(2)$ corresponding three-dimensional
rotation group.  If all the elements are real numbers, the group becomes
the three-parameter $Sp(2)$ group.  This subgroup is  equivalent to
$SU(1,1)$ which is the primary scientific language for squeezed states
of light~\cite{yuen76,knp91}.

We can also consider the matrix of Eq.(\ref{alpha}) when one of its
off-diagonal elements vanishes.  Then, it takes the form
\begin{equation}
   \pmatrix{\exp{(i\phi/2)}   &    0  \cr \gamma &
   \exp{(-i\phi/2)} } ,
\end{equation}
where $\gamma$ is a complex number with two real parameters.  In
1939~\cite{wig39},
Wigner observed this form as one of the subgroups of the Lorentz group.
He observed further that this group is isomorphic to the two-dimensional
Euclidean group, and that its four-by-four equivalent can explain the
internal space-time symmetries of massless particles including the photons.

In ray optics, we often have to deal with this type of triangluar matrices,
particularly in lens optics and stability problems in laser  and
multilayer optics.  In the language of mathemtics, dealing with this form
is called the Iwasawa decomposition~\cite{iwa49}.  This aspect of the
Lorentz group is dicussed in Appendix~\ref{conju}.

\section{Polarization Optics and Interferometers}\label{polari}
In polarization physics, we are studying optical rays with
two independent components.  Thus the two-by-two matrix
given in Sec~\ref{intro} is directly applicable.
We are quite familiar with Pauli spinors and Pauli
matrices with three independent parameters. They deal with
rotations.  If we add Lorentz boosts, the group becomes
the six-parameter $SL(2,c)$ group.  The representation is
still two-by-two.

 From the group theoretical point of view, the Jones matrices
constitute the $SL(2,c)$ group or the universal covering
group for the Lorentz group.  The Jones vectors and Jones
matrices are nothing more and less than the $SL(2, C)$
representation and the $SL(2)$ spinors.

There is a specific procedure to construct the Minkowskian
four-vectors and the four-by-four Lorentz transformation
matrices. Then they become Stokes parameters and the
Mueller matrices.

If the Jones matrix contains all the parameters for
the polarized light beam, why do we need the mathematics in
the four-dimensional space?  The answer to this question is
well known.  In addition to the basic parameter given by the
Jones vector, the Stokes parameters give the degree of
coherence between the two rays.

Let us write Eq.(\ref{jones}) as  a Jones spinor of the form
\begin{equation}
 \pmatrix{\psi_{1}(z, t) \cr \psi_{2}(z, t) } ,
\end{equation}
and  construct the quantities:
\begin{eqnarray}
&{}&  S_{11} = <\psi_{1}^{*}\psi_{1}> , \qquad
  S_{12} = <\psi_{1}^{*}\psi_{2}> , \nonumber \\[2ex]
&{}&  S_{21} = <\psi_{2}^{*}\psi_{1}> , \qquad
  S_{22} = <\psi_{2}^{*}\psi_{2}>  .
\end{eqnarray}
Then the Stokes vector consists of
\begin{eqnarray}
&{}&  S_{0} = S_{11} + S_{22} , \quad
  S_{1} = S_{11} - S_{22}  , \nonumber \\[2ex]
&{}&  S_{2} = S_{12} + S_{21}  , \quad
  S_{3} = -i(S_{11} + S_{22})   .
\end{eqnarray}
The four-component vector $(S_{0}, S_{1}, S_{2}, S_{3})$
transforms like the space-time four-vector $(t, z, x, y)$ under
Lorentz transformations.
The Mueller matrix is therefore like the Lorentz-transformation
matrix.

As in the case of special relativity,
let us consider the quantity
\begin{equation}
M^{2} = S_{0}^{2}- S_{1}^{2} - S_{2}^{2} - S_{3}^{2} .
\end{equation}
Then $M$ is like the mass of the particle while the Stokes four-vector
is like the four-momentum.

If $M = 0$,  the two-beams are in a purely state. As $M$ increases,
the system becomes mixed, and the entropy increases. If it reaches the
value of $S_{0}$, the system becomes completely random.  It is
gratifying to note that this mechanism can be formulated in terms of
the four-momentum in particle physics~\cite{hkn99}.

\section{One-lens System}\label{olens}
In analyzing optical rays in para-axial lens optics, we start with
the lens matrix:
\begin{equation}\label{lens}
 L = \pmatrix{1 & 0 \cr -1/f & 1} ,
\end{equation}
and the translation matrix
\begin{equation}\label{trans}
T = \pmatrix{1 & z \cr 0 & 1} ,
\end{equation}
assuming that the beam is propagating along the $Z$ direction.

Then the one-lens system consists of
\begin{equation}
\pmatrix{1 & z_{2} \cr 0 & 1}
\pmatrix{1 & 0 \cr -1/f & 1}
\pmatrix{1 & z_{1} \cr 0 & 1} .
\end{equation}
If we perform the matrix multiplication,
\begin{equation}
\pmatrix{1 - z_{2}/f   & z_{1} + z_{2} - z_{1}z_{2}/f   \cr
-1/f & 1 - z_{1}/f } .
\end{equation}
If we assert that the upper-right element be zero,  then
\begin{equation}
{1 \over z_{1}} + {1 \over z_{2}} = {1 \over f} ,
\end{equation}
and the image is focussed, where $z_{1}$ and $z_{2}$ are the distance
between the lens and object and between the lens and image respectively.
They are in general different, but we shall assume for simplicity that
they are the same: $z_{1} = z_{2} = z$.  We are doing this because this
simplicity does not
destroy the main point of our discussion, and because the case with
two different values has been dealt with in the literature~\cite{gk03}.
Under this assumption, we are left with
\begin{equation}
\pmatrix{1 - z/f  &  2z - z^{2}/f  \cr  -1/f  & 1 - z/f } .
\end{equation}
The diagonal elements of this matrix are dimensionless.  In order to
make the off-diagonal elements dimensionless, we write this matrix
as
\begin{equation}
 - \pmatrix{\sqrt{z}  & 0 \cr 0 & 1/\sqrt{z}}
\pmatrix{1 - z/f &  z/f - 2  \cr  z/f  & 1 - z/f}
\pmatrix{\sqrt{z}  & 0 \cr 0 & 1/\sqrt{z}} .
\end{equation}
Indeed, the matrix in the middle contains dimensionless elements.  The
negative sign in front is purely for convenience.  We are then led to
study the core matrix
\begin{equation}\label{core}
C = \pmatrix{ x - 1  & x -2  \cr x  & x - 1} .
\end{equation}

Here, the important point is that the above matrices can be written in
terms of transformations in the Lorentz group.  In the two-by-two matrix
representation, the Lorentz boost along the $z$ direction takes the form
\begin{equation}
B(\eta) = \pmatrix{\exp{(\eta/2)} & 0 \cr 0 & \exp{(-\eta/2)} } ,
\end{equation}
and the rotation around the $y$ axis can be written as
\begin{equation}
R(\theta) = \pmatrix{\cos(\phi/2) & -\sin(\phi/2) \cr
\sin(\phi/2) & \cos(\phi/2) } ,
\end{equation}
and the boost along the $x$ axis takes the form
\begin{equation}
X(\chi) = \pmatrix{\cosh(\chi/2) & \sinh(\chi/2) \cr
\sinh(\chi/2) & \cosh(\chi/2)} .
\end{equation}
Then the core matrix of Eq.(\ref{core}) can be written as
\begin{equation}\label{phi11}
B(\eta) R(\phi) B(-\eta),
\end{equation}
or
\begin{equation}\label{phi22}
\pmatrix{\cos(\phi/2)  &
- e^{-\eta}\sin(\phi/2) \cr  e^{\eta}\sin(\phi/2) & \cos(\phi/2) } ,
\end{equation}
if $ 1 < x < 2 $.  If $x$ is greater than 2, the upper-right element
of the core is positive and it can take the form
\begin{equation}\label{chi11}
B(\eta) X(\chi) B(-\eta),
\end{equation}
or
\begin{equation}\label{chi22}
 \pmatrix{\cosh(\chi/2)  &
 e^{-\eta}\sinh(\chi/2) \cr  e^{+\eta}\sinh(\chi/2) & \cosh(\chi/2) } .
\end{equation}

The expressions of Eq.(\ref{phi11}) and Eq.(\ref{chi11}) are a Lorentz
boosted rotation and a Lorentz-boosted boost matrix along the $x$
direction respectively.  These expressions play the key role in
understanding Wigner's little groups for relativistic particles.

Let us look at their explicit matrix representations given
in Eq.(\ref{phi22}) and Eq.(\ref{chi22}).  The transition from
Eq.(\ref{phi22}) to Eq.(\ref{chi22}) requires the upper right element
going through zero.  This can only be achieved through $\eta$
going to infinity.  If we like to keep the lower-left element finite
during this process, the angle $\phi$ and the boost parameter $\chi$
have to approach zero.  The process of approaching the vanishing
upper-right element is necessarily a singular transformation.
This aspect plays the key role in unifying the internal space-time
symmetries of massive and massless particles.  This is like
Einstein's $E = \sqrt{(pc)^{2} + m_{0}^{2}c^{4}}$ becoming
$E = pc$ in the limit of large momentum.

On the other hand, the core matrix of Eq.(\ref{core}) is an analytic
function of the variable $x$.  Thus, the lens matrix allows a
parametrization which allows the transition from massive particle to
massless particle analytically.  The lens optics indeed serves as the
analogue computer for this important transition in particle physics.

>From the mathematical point of view,  Eq.(\ref{phi22}) and Eq.(\ref{chi22})
represent circular and hyperbolic geometries, respectively.  The
transition from one to the other is not a trivial mathematical procedure.
It requires a further investigation.

Let us go back to the core matrix of Eq.(\ref{core}).  The $x$ parameter
does not appear to be a parameter of Lorentz transformations.  However,
the matrix can be written in terms of another set of Lorentz transformations.
This aspect has been discussed in the literature~\cite{bk03}.

\section{Multi-lens Problem}\label{mlens}
Let us consider a co-axial system of an arbitrary number of lens.
Their focal lengths are not necessarily the same, nor are their separations.
We are then led to consider an arbitrary number of the lens matrix given
in Eq.(\ref{lens}) and an arbitrary number of translation matrix
of Eq.(\ref{trans}).  They are multiplied like
\begin{equation}\label{lsystem}
T_{1}~ L_{1}~T_{2}~ L_{2}~T_{3}~ L_{3}............T_{N}~ L_{N} ,
\end{equation}
where $N$ is the number of lenses.

The easiest way to tackle this problem in to use the Lie-algebra
approach.  Let us start with the generators of the Sp(2) group:
\begin{eqnarray}\label{sq11}
B_{1} = { 1 \over 2}\pmatrix{i  & 0 \cr 0 & -i }, \qquad
B_{2} = { 1 \over 2}\pmatrix{0  & i \cr i & 0 }, \qquad
J = { 1 \over 2}\pmatrix{0  & -i \cr i & 0 } .
\end{eqnarray}
Since the generators are pure imaginary, the transformation matrices
are real.

On the other hand,  the $L$ and $T$ matrices of Eq.(\ref{lens}) and
Eq.(\ref{trans}) are generated by
\begin{equation}
X_{1} = \pmatrix{ 0 & i \cr 0 & 0},  \qquad
X_{2} = \pmatrix{ 0 & 0 \cr i & 0} .
\end{equation}
If we introduce the third matrix
\begin{equation}
X_{3} = \pmatrix{i & 0 \cr 0 & -i} ,
\end{equation}
all three matrices form a closed set of commutation relations:
\begin{equation}\label{shear}
\left[X_{1}, X_{2}\right] = iX_{3}, \qquad
\left[X_{1}, X_{3}\right] = -iX_{1},  \qquad
\left[X_{2}, X_{3}\right] = iX_{2} .
\end{equation}
Thus, these generators also form a closed set of Lie algebra
generating real two-by-two matrices.  What group would this
generate?  The answer has to be $Sp(2)$.  The truth is that
the three generators given in Eq.(\ref{shear}) can be written as
linear combinations of the generators of the $Sp(2)$ group given
in Eq.(\ref{sq11})~\cite{bk01}.  Thus, the $X_{i}$ matrices
given above can also act as the generators of the $Sp(2)$ group,
and the lens-system matrix given in Eq.(\ref{lsystem}) is a
three-parameter matrix of the form of Eq.(\ref{alpha}) with
real elements.

The resulting real matrix is written as
\begin{equation}
\pmatrix{A & B \cr C & D} ,
\end{equation}
and is called the $ABCD$ matrix.  According to Bargamann~\cite{barg47},
this three-parameter matrix can be decomposed into
\begin{equation}
\pmatrix{\cos(\alpha/2)  & -\sin(\alpha/2)
\cr \sin(\alpha/2)  & \cos(\alpha/2)}
\pmatrix{e^{\gamma/2} & 0 \cr  0 & e^{-\gamma/2} }
\pmatrix{\cos(\beta/2)  & -\sin(\beta/2)  \cr
\sin(\beta/2)  & \cos(\beta/2)},
\end{equation}
which can be written as the product of one symmetric matrix resulting
from
\begin{equation}
\pmatrix{\cos(\alpha/2)  & -\sin(\alpha/2)
\cr \sin(\alpha/2)  & \cos(\alpha/2)}
\pmatrix{e^{\gamma/2} & 0 \cr  0 & e^{-\gamma/2} }
\pmatrix{\cos(\alpha/2)  & -\sin(\alpha/2)  \cr
\sin(\alpha/2)  & \cos(\alpha/2)}
\end{equation}
and one rotation matrix:
\begin{equation}
\pmatrix{\cos[(\beta - \alpha)/2]  & -\sin[(\beta - \alpha)/2]   \cr
\sin[(\beta - \alpha)/2]   & \cos[(\beta - \alpha)/2] } .
\end{equation}
We can then decompose each of these two matrices
into the lens and translation matrices.  The net result is that
we do not need more than three lenses to describe the lens system
consisting of an arbitrary number of lenses.  The detailed
calculations are given in Ref.~\cite{bk01}.

\section{Laser Cavities}\label{lcav}
In a laser cavity, the optical ray makes round trips between two mirrors.
One cycle is therefore equivalent to a two-lens system with
two identical lenses and the same  distance between the lenses.  Let us
rewrite the matrix corresponding to the one-lens system given in
Eq.(\ref{core}).
\begin{equation}\label{core22}
C = \pmatrix{ x - 1  & x -2  \cr x  & x - 1} .
\end{equation}
Then one complete cycle consists of $C^{2}$.  For $N$cycles, the
expression should be $C^{2N}$.  However this calculation, using the
above expression, will not lead to a manageable form.  However, we
can resort to the expressions of Eq.(\ref{phi11}) and Eq.(\ref{chi11}).
Then one cycle consists of
\begin{eqnarray}\label{cycle}
&{}& C^{2} = [B(\eta) R(\phi/2) B(-\eta)]
[B(\eta) R(\phi/2) B(-\eta)]    \nonumber \\[2ex]
&{}& \hspace{5mm} = B(\eta) R(\phi) B(-\eta) ,
\end{eqnarray}
if the upper-right element is negative.  If it is positive, the
expression should be
\begin{eqnarray}
&{}& C^{2} = [B(\eta) X(\chi/2) B(-\eta)]
[B(\eta) X(\chi/2) B(-\eta)]    \nonumber \\[2ex]
&{}& \hspace{5mm} = B(\eta) X(\chi) B(-\eta) .
\end{eqnarray}

If these expressions are repeated $N$ times,
\begin{equation}\label{stable}
C^{2N} = B(\eta) R(N\phi) B(-\eta) ,
\end{equation}
if the upper-right element is negative.  It it is positive, the
expression should be
\begin{equation}
C^{2N} = B(\eta) X(N\chi) B(-\eta) .
\end{equation}
As $N$ becomes large,  $\cosh(N\chi)$ and $\sinh(N\chi)$ become
very large, the beam deviates from the laser cavity.   Thus,
we have to restrict ourselves to the case given in Eq.(\ref{stable}).

The core of the expression of Eq.(\ref{stable}) is the rotation
matrix
\begin{equation}
       R(N\phi) = [R(\phi)]^{N}     .
\end{equation}
This means that one complete cycle in the cavity corresponds to
the rotation matrix $R(\phi)$.  The rotation continues as the
beam continues to repeat the cycle.

Let us go back to Eq.(\ref{cycle}).  The expression corresponds to
a Lorentz boosted rotation, or bringing a moving particle to its rest
frame, rotate, and boost back to the original momentum.  The rotation
associated with the momentum-preserving transformation is called
the Wigner's little-group rotation, which is related to the
Wigner rotation commonly mentioned in the literature~\cite{bk01}.

\section{Multilayer Optics}\label{mlayer}
The most efficient way to study multilayer optics is to use the S-matrix
formalism~\cite{azzam77}.  We consider a system of two different
optical layers.  For convenience,
we start from the boundary from $medium~2$ to $medium~1$.
We can write the boundary matrix as~\cite{monzon00}
\begin{equation}\label{bd11}
B(\eta) = \pmatrix{\cosh(\eta/2) & \sinh(\eta/2) \cr \sinh(\eta/2) &
\cosh(\eta/2) } ,
\end{equation}
taking into account both the transmission and reflection of the beam.  As
the beam goes through the $medium~1$, the beam undergoes the phase
shift represented by the matrix
\begin{equation}\label{ps11}
P(\phi_1) = \pmatrix{e^{-i\phi_1/2} & 0 \cr 0 & e^{i\phi_1/2}} .
\end{equation}
When the wave hits the surface of the second medium, the boundary
matrix is
\begin{equation}\label{bd22}
B(-\eta) = \pmatrix{\cosh(\eta/2) & -\sinh(\eta/2) \cr -\sinh(\eta/2) &
\cosh(\eta/2) } ,
\end{equation}
which is the inverse of the matrix given in Eq.(\ref{bd11}).
Within the second medium, we write the phase-shift matrix as
\begin{equation}\label{ps22}
P(\phi_2) = \pmatrix{e^{-i\phi_2/2} & 0 \cr 0 & e^{i\phi_2/2}} .
\end{equation}
Then, when the wave hits the first medium from the second, we have
to go back to Eq.(\ref{bd11}).
Thus, the one cycle consists of
\begin{eqnarray}\label{m1}
&{}& \pmatrix{\cosh(\eta/2) &
\sinh(\eta/2) \cr \sinh(\eta/2) & \cosh(\eta/2) }
\pmatrix{e^{-i\phi_1/2} & 0 \cr 0 & e^{i\phi_1/2}}\nonumber \\[2ex]
&{}& \hspace{10mm}
\times \pmatrix{\cosh(\eta/2) & -\sinh(\eta/2)
\cr -\sinh(\eta/2) & \cosh(\eta/2) }
\pmatrix{e^{-i\phi_2/2} & 0 \cr 0 & e^{i\phi_2/2}} .
\end{eqnarray}

The above matrices contain complex numbers.  However, it is possible
to transform simultaneously
\begin{equation}
\pmatrix{\cosh(\eta/2) & \sinh(\eta/2) \cr \sinh(\eta/2) &  \cosh(\eta/2)}
\end{equation}
to
\begin{equation}
\pmatrix{\exp{(\eta/2)} & 0 \cr 0 & \exp{(-\eta/2)} },
\end{equation}
and transform
\begin{equation}
\pmatrix{e^{-i\phi_i/2} & 0 \cr 0 & e^{i\phi_i/2}}
\end{equation}
to
\begin{equation}
\pmatrix{\cos(\phi_i/2) & -\sin(\phi_i/2) \cr
\sin(\phi_i/2)  & \cos(\phi_i/2) } ,
\end{equation}
using a conjugate transformation.  It is also possible to transform
these expressions back to their original forms.  This transformation
property has been discussed in detail in Ref.~\cite{gk01}.

As a consequence, the matrix of Eq.(\ref{m1}) becomes
\begin{eqnarray}\label{core55}
&{}&\pmatrix{e^{\eta/2} & 0 \cr 0 & e^{-\eta/2} }
\pmatrix{\cos(\phi_1/2) & -\sin(\phi_1/2) \cr
\sin(\phi_1/2) &  \cos(\phi_1/2)}   \nonumber \\[2ex]
&{}&  \hspace{10mm}\times \pmatrix{ e^{-\eta/2} & 0 \cr
0 & e^{\eta/2}}
\pmatrix{\cos(\phi_2/2) & -\sin(\phi_2/2) \cr
\sin(\phi_2/2) &  \cos(\phi_2/2) }  .
\end{eqnarray}
In the above expression, the first three matrices are of the
same mathematical form as that of core matrix for the one-lens system
given in Eq.(\ref{phi11}).  The fourth matrix is an additional rotation
matrix.  This makes the mathematics of repetition more complicated, but
this has been done~\cite{gk03}.

As a consequence the net result becomes
\begin{equation}
      B(\mu) R(N\alpha) B(-\mu) ,
\end{equation}
or
\begin{equation}
      B(\mu) X(N\xi) B(-\mu),
\end{equation}
where the parameters $\mu, \alpha$ and $\xi$ are to be determined from
the input parameters $\eta, \phi_1$ and $\phi_2$.  Detailed calculations
are given in Ref.~\cite{gk03}.

It is interesting to note that the Lorentz group can serve as
a computational device also in multilayer optics.

\section{Concluding Remarks}
We have seen in this report that the Lorentz group provides convenient
calculational tools in many branches of ray optics.  The reason is that
ray optics is largely based on two-by-two matrices.   These matrices
also constitute the group $SL(2,c)$ which serves as the universal
covering group of the Lorentz group.

The optical instruments discussed in this report are the fundamental
components in optical circuits.  In the world of electronics, electric
circuits form the fabric of the system.  In the future high-technology
world, optical components will hold the key to technological
advances.  Indeed, the Lorentz group is the fundamental language for
the new world.

It is by now well known that the Lorentz group is the basic language
for quantum optics.  Coherent and squeezed states are representations
of the Lorentz group. It is challenging to see how the Lorentz nature
of the above-mentioned optical components will manifest itself in
quantum world.

The Lorentz group was introduced to physics by Einstein and Wigner to
understand the space-time symmetries of relativistic particles and
the covariant world of electromagnetic fields. It is gratifying to note
that the Lorentz group can serve as the language common both to
particle physics and optical sciences.

\section*{Appendix}

\appendix

\section{Lorentz Transformations}\label{lorentz}
Let us consider the space-time coordinates $(t, z, x, y)$.  Then the
rotation around the $z$ axis is performed by the four-by-four matrix
\begin{equation}
\pmatrix{1 & 0 & 0 & 0 \cr 0 & 0 & 0 & 0 \cr
0 & 0 & \cos\theta & -\sin\theta \cr 0 & 0 & \sin\theta & \cos\theta} .
\end{equation}
This transformation is generated by
\begin{equation}
J_{3} = \pmatrix{0 & 0 & 0 & 0 \cr 0 & 0 & 0 & 0 \cr
0 & 0 & 0 & -i \cr 0 & 0 & i & 0} .
\end{equation}
Likewise, we can write down the generators of rotations $J_{1}$ and
$J_{2}$ around the $x$ and $y$ axes respectively.
\begin{equation}
J_{1} = \pmatrix{0 & 0 & 0 & 0 \cr 0 & 0 & 0 & i \cr
0 & 0 & 0 & 0 \cr 0 & -i & 0 & 0} , \qquad
J_{2} = \pmatrix{0 & 0 & 0 & 0 \cr 0 & 0 & -i & 0 \cr
0 & i & 0 & 0 \cr 0 & 0 & 0 & 0} .
\end{equation}
These three generators
satisfy the closed set of commutations relations
\begin{equation}\label{rota}
\left[J_{i}, J_{j} \right] = i \epsilon_{ijk} J_{k} .
\end{equation}
This set of commutation relations is for the three-dimensional rotation
group.

The Lorentz boost along the $z$ axis takes the form
\begin{equation}
\pmatrix{\cosh\eta & \sinh\eta & 0 & 0 \cr
\sinh\eta & \cosh\eta & 0 & 0 \cr
0 & 0 & 1 & 0 \cr 0 & 0 & 0 & 1} ,
\end{equation}
which is generated by
\begin{equation}
K_{3} = \pmatrix{0 & i & 0 & 0 \cr i & 0 & 0 & 0 \cr
0 & 0 & 0 & 0 \cr 0 & 0 & 0 & 0} .
\end{equation}
Likewise, we can write generators of boosts $K_{1}$ and $K_{2}$ along the
$x$ and $y$ axes respectively, and they take the form
\begin{equation}
K_{1} = \pmatrix{0 & 0 & i & 0 \cr 0 & 0 & 0 & 0 \cr
i & 0 & 0 & 0 \cr 0 & 0 & 0 & 0} , \qquad
K_{2} = \pmatrix{0 & 0 & 0  & i \cr 0 & 0 & 0 & 0 \cr
0 & 0 & 0 & 0 \cr i & 0 & 0 & 0} .
\end{equation}
These boost generators satisfy the commutation relations
\begin{equation}\label{boosta}
\left[J_{i}, K_{j} \right] = i \epsilon_{ijk} K_{k} , \qquad
\left[K_{i}, K_{j} \right] = -i \epsilon_{ijk} J_{k} .
\end{equation}

Indeed, the three rotation generators and the three boost generators
satisfy the closed set of commutation relations given in Eq.(\ref{rota})
and Eq.(\ref{boosta}).  These three commutation relations form the
starting point of the Lorentz group.  The generators given in this
Appendix are four-by-four matrices, but they are not the only set
satisfying the commutation relations.   We can construct also six
two-by-two matrices satisfying the same set of commutation relations.
The group of transformations constructed from these two-by-matrices
is often called $SL(2,c)$ or the two-dimensional representation of
the Lorentz group.
Throughout the present paper, we used the two-by-two transformation
matrices constructed from the generators of the $SL(2,c)$ group.

\section{Spinors and Four-vectors in the Lorentz Group}\label{spinor}
In Appendix A, we have noted that there are four-by-four and two-by-two
representations of the Lorentz group.  The four-by-four representation
is applicable to covariant four-vectors, while the two-by-two
transformation matrices are applicable to two-component spinors which
in the present case are Jones vectors.  The question then is whether
we can construct the four-vector from the spinors.  In the language of
polarization optics, the question is whether it is possible to
construct the coherency matrix~\cite{born80,perina71} from the Jones
vector.

With this point in mind, let us start from the following form of the
Pauli spin matrices.
\begin{equation}
\sigma_{1} = \pmatrix{1 & 0 \cr 0 & -1} , \quad
\sigma_{2} = \pmatrix{0 & 1 \cr 1 & 0} , \quad
\sigma_{3} = \pmatrix{0 & -i \cr i & 0} .
\end{equation}
These matrices are written in a different convention.  Here  $\sigma_{3}$
is imaginary, while $\sigma_{2}$ is imaginary in the traditional notation.
Also in this convention, we can construct three rotation generators
\begin{equation}
J_{i} = {1 \over 2} \sigma_{i} ,
\end{equation}
which satisfy the closed set of commutation relations
\begin{equation}\label{comm1}
\left[J_{i}, J_{j}\right] = i \epsilon_{ijk} J_{k} .
\end{equation}
We can also construct three boost generators
\begin{equation}
K_{i} = {i \over 2} \sigma_{i} ,
\end{equation}
which satisfy the commutation relations
\begin{equation}\label{comm2}
\left[K_{i}, K_{j}\right] = -i \epsilon_{ijk} J_{k} .
\end{equation}
The $K_{i}$ matrices alone do not form a closed set of commutation
relations, and the rotation generators $J_{i}$ are needed to form a
closed set:
\begin{equation}\label{comm3}
\left[J_{i}, K_{j}\right] = i \epsilon_{ijk} K_{k} .
\end{equation}

The six matrices $J_{i}$ and $K_{i}$ form a closed set of commutation
relations, and they are like the generators of the Lorentz group applicable
to the (3 + 1)-dimensional Minkowski space.  The group generated by the
above six matrices is called $SL(2,c)$ consisting of all two-by-two complex
matrices with unit determinant.

In order to construct four-vectors, we need two different spinor
representations of the Lorentz group.  Let us go to the commutation
relations for the generators given in Eqs.(\ref{comm1}), (\ref{comm2}) and
(\ref{comm3}).  These commutators are
not invariant under the sign change of the rotation generators $J_{i}$,
but are invariant under the sign change of the squeeze operators $K_{i}$.
Thus, to each spinor representation, there is another representation with
the squeeze generators with opposite sign.  This allows us to construct
another representation with the generators:
\begin{equation}
\dot{J}_{i} = {1 \over 2} \sigma_{i}, \qquad
\dot{K}_{i} = {-i \over 2} \sigma_{i} .
\end{equation}
We call this representation the ``dotted'' representation.  If we write
the transformation matrix $L$ of Eq.(\ref{alpha}) in terms of the
generators as
\begin{equation}
L = \exp\left\{-{i\over 2} \sum_{i=1}^{3}\left(\theta_{i}\sigma_{i} +
i\eta_{i}\sigma_{i}\right) \right\} ,
\end{equation}
then the transformation matrix in the dotted representation becomes
\begin{equation}\label{eldot}
\dot{L} = \exp\left\{-{i\over 2} \sum_{i=1}^{3}\left(\theta_{i}\sigma_{i}
- i\eta_{i}\sigma_{i}\right)\right\} .
\end{equation}
In both of the above matrices, Hermitian conjugation changes the
direction of rotation.  However, it does not change the direction of
boosts.  We can achieve this only by interchanging $L$ to $\dot{L}$,
and we shall call this the ``dot'' conjugation.

Likewise, there are two different set of spinors.  Let us use $u$ and
$v$ for the up and down spinors for ``undotted'' representation.  Then
$\dot{u}$ and $\dot{v}$ are for the dotted representation.  The
four-vectors are then constructed as~\cite{hks86}
\begin{eqnarray}
&{}&  u\dot{u} = - (x - iy), \quad v\dot{v} = (x + iy), \nonumber \\[2ex]
&{}&   u\dot{v} = (t + z), \quad v\dot{u} = -(t - z)
\end{eqnarray}
leading to the matrix
\begin{equation}\label{dotmat}
C = \pmatrix{u \dot{v} & -u\dot{u} \cr v\dot{v} & -v\dot{u}}
   = \pmatrix{u \cr v} \pmatrix {\dot{v} & -\dot{u}} ,
\end{equation}
where $u$ and $\dot{u}$ are one if the spin is up, and are zero if the
spin is down, while $v$ and $\dot{v}$ are zero and one for the spin-up
and spin-down cases.
The transformation matrix applicable to the column vector in the above
expression is the two-by-two matrix given in Eq.(\ref{alpha}).  What
is then the transformation matrix applicable to the row vector
$(\dot{v},~-\dot{u})$ from the right-hand side?  It is the transpose
of the matrix applicable to the column vector $(\dot{v},~-\dot{u})$.
We can obtain this column vector from
\begin{equation}\label{dotcol}
 \pmatrix {\dot{v} \cr -\dot{u}} ,
\end{equation}
by applying to it the matrix
\begin{equation}
g = -i\sigma_{3} = \pmatrix{0 & -1 \cr 1 & 0} .
\end{equation}
This matrix also has the property
\begin{equation}
g \sigma_{i} g^{-1} = -\left(\sigma_{i}\right)^{T} ,
\end{equation}
where the superscript $T$ means the transpose of the matrix.  The
transformation matrix applicable to the column vector of Eq.(\ref{dotcol})
is $\dot{L}$ of Eq.(\ref{eldot}).  Thus the matrix applicable to the row
vector $(\dot{v},~-\dot{u})$ in Eq.(\ref{dotmat}) is
\begin{equation}
\left\{g^{-1} \dot{L} g\right\}^{T} = g^{-1} \dot{L}^{T} g .
\end{equation}
This is precisely the Hermitian conjugate of $L$.

In optics, this two-by-two matrix form appears as the coherency matrix,
and it takes the form
\begin{equation}\label{cohm1}
C = \pmatrix{<E^{*}_{x}E_{x}> & <E^{*}_{y} E_{x}> \cr
<E^{*}_{x} E_{y}> & <E^{*}_{y} E_{y}>} ,
\end{equation}
where $<E^{*}_{i}E_{j}>$ is the time average of $E^{*}_{i}E_{j}$.
This matrix is convenient when we deal with light waves whose two transverse
components are only partially coherent.  In terms of the complex parameter
$w$, the coherency matrix is proportional to
\begin{equation}\label{cohm2}
C = \pmatrix{1 & r e^{i\delta} \cr
r e^{-i\delta} & r^{2}} ,
\end{equation}
if the $x$ and $y$ components are perfectly coherent with the phase
difference of $\delta$.  If they are totally incoherent, the
off-diagonal elements vanish in the above matrix.

Let us now consider its transformation properties.  As was noted by
Opatrny and Perina~\cite{perina93}, the matrix of Eq.(\ref{cohm1}) is
like
\begin{equation}\label{matC}
C = \pmatrix{t + z & x - iy \cr x + iy & t - z} ,
\end{equation}
where the set of variables $(x, y, z, t)$ is transformed like a four-vector
under Lorentz transformations.  Furthermore, it is known that the Lorentz
transformation of this four-vector is achieved through the formula
\begin{equation}\label{Ldag}
C' = L C L^{\dagger} ,
\end{equation}
where the transformation matrix $L$ is that of Eq.(\ref{alpha}).  The
construction of four-vectors from the two-component spinors is not
a trivial task~\cite{hks86,bask95}.  The two-by-two representation of
Eq.(\ref{matC}) requires one more step of complication.

\section{Conjugate Transformations}\label{conju}
The core matrix of Eq.(\ref{core}) contains the chain of the matrices
\begin{equation}\label{su11}
W =\pmatrix{e^{-i\phi} & 0 \cr 0 & e^{i\phi}}
\pmatrix{\cosh\eta & \sinh\eta \cr \sinh\eta & \cosh\eta}
\pmatrix{e^{-i\xi} & 0 \cr 0 & e^{i\xi}}  .
\end{equation}
The Lorentz group allows us to simplify this expression under
certain conditions.

For this purpose, we transform the above expression into a more
convenient form, by taking the conjugate of each of the matrices with
\begin{equation}
C_{1} =  {1 \over \sqrt{2}} \pmatrix{1 & i \cr i & 1} .
\end{equation}
Then $C_{1} W C_{1}^{-1}$ leads to
\begin{equation}\label{sp2}
\pmatrix{\cos\phi & -\sin\phi \cr \sin\phi & \cos\phi}
\pmatrix{\cosh\eta & \sinh\eta \cr \sinh\eta & \cosh\eta}
\pmatrix{\cos\xi & -\sin\xi \cr \sin\xi & \cos\xi} .
\end{equation}
In this way, we have converted $W$ of Eq.(\ref{su11}) into a real
matrix, but it is not simple enough.

Let us take another conjugate with
\begin{equation}
C_{2} =  {1 \over \sqrt{2}} \pmatrix{1 & 1 \cr -1 & 1} .
\end{equation}
Then the conjugate $C_{2} C_{1} W C_{1}^{-1} C_{2}^{-1} $ becomes
\begin{equation}\label{abcd}
\pmatrix{\cos\phi & -\sin\phi \cr \sin\phi & \cos\phi}
\pmatrix{e^{\eta} &  0 \cr 0 & e^{-\eta}}
\pmatrix{\cos\xi & -\sin\xi \cr \sin\xi & \cos\xi} .
\end{equation}
The combined effect of $C_{2}C_{1}$ is
\begin{equation}\label{ccc}
C = C_{2}C_{1} = {1 \over \sqrt{2}} \pmatrix{e^{i\pi/4} &  e^{i\pi/4} \cr
-e^{-i\pi/4} & e^{-i\pi/4}} ,
\end{equation}
with
\begin{equation}
C^{-1} = {1 \over \sqrt{2}} \pmatrix{e^{-i\pi/4} &  -e^{i\pi/4} \cr
e^{-i\pi/4} & e^{i\pi/4}} .
\end{equation}

After multiplication, the matrix of Eq.(\ref{abcd}) will take the form
\begin{equation}
V = \pmatrix{A &  B \cr C & D} ,
\end{equation}
where $A, B, C,$ and $D$ are real numbers.  If $B$ and $C$ vanish, this
matrix will become diagonal, and the problem will become too simple.
If, on the other hand, only one of these two elements become zero, we will
achieve a substantial mathematical simplification and will be encouraged
to look for physical circumstances which will lead to this simplification.

Let us summarize.  we started in this section with the matrix
representation $W$ given in Eq.(\ref{su11}).  This form can be transformed
into the $V$ matrix of Eq.(\ref{abcd}) through the conjugate
transformation
\begin{equation}\label{conju11}
V = C W C^{-1} ,
\end{equation}
where $C$ is given in Eq.(\ref{ccc}).  Conversely, we can recover
the $W$ representation by
\begin{equation}\label{conj22}
W = C^{-1} V C .
\end{equation}
For calculational purposes, the $V$ representation is much easier
because we are dealing with real numbers.  On the other hand, the
$W$ representation is of the form for the S-matrix we intend to compute.
It is gratifying to see that they are equivalent.

Let us go back to Eq.(\ref{abcd}) and consider the case where
the angles $\phi$ and $\xi$ satisfy the following constraints.
\begin{equation}\label{2angs}
\phi + \xi = 2\theta, \qquad \phi - \xi = \pi/2 ,
\end{equation}
thus
\begin{equation}\label{phixi}
\phi =  \theta + \pi/4,  \qquad   \xi = \theta - \pi/4 .
\end{equation}
Then in terms of $\theta$, we can reduce the matrix of Eq.(\ref{abcd})
to the form
\begin{equation}\label{abcd2}
\pmatrix{(\cosh\eta)\cos(2\theta)  &
       \sinh\eta - (\cosh\eta)\sin(2\theta)  \cr
       \sinh\eta + (\cosh\eta)\sin(2\theta)  &
(\cosh\eta)\cos(2\theta) } .
\end{equation}
Thus the matrix takes a surprisingly simple form if the parameters
$\theta$ and $\eta$ satisfy the constraint
\begin{equation}\label{constr}
\sinh\eta = (\cosh\eta)\sin(2\theta) .
\end{equation}\
Then the matrix becomes
\begin{equation}\label{decom2}
\pmatrix{1  &  0  \cr 2\sinh\eta & 1 }  .
\end{equation}
This aspect of the Lorentz group is known as the Iwasawa
decomposition~\cite{iwa49}, and has been discussed in the optics
literature~\cite{simon98,hkn99}.

The matrices of the form is not so strange in optics.  In para-axial
lens optics, the translation and lens matrices are written as
\begin{equation}\label{shear1}
\pmatrix{1 & u \cr 0 & 1} , \quad and  \quad \pmatrix{1 & 0 \cr u & 1} ,
\end{equation}
respectively.  These matrices have the following interesting mathematical
property~\cite{hkn97},
\begin{equation}
\pmatrix{1 & u_{1} \cr 0 & 1} \pmatrix{1 & u_{2} \cr 0 & 1} =
\pmatrix{1 & u_{1} + u_{2} \cr 0 & 1} ,
\end{equation}
and
\begin{equation}
\pmatrix{1 & 0 \cr u_{1} & 1} \pmatrix{1 & 0 \cr u_{1} & 1}
= \pmatrix{1 & 0 \cr u_{1} + u_{2} & 1} .
\end{equation}
We note that the multiplication is commutative, and the parameter
becomes additive.  These matrices convert multiplication into
addition, as logarithmic functions do.

\section{Euler versus Lie Representations}\label{euler}
In this paper, we restricted ourselves to the algebra of two-by-two
matrices and avoided as much as possible group theoretical languages.
In order to explain where those algebraic tricks came from, we give
in this Appendix A group theoretical interpretation of what we did
in this paper.

The group $SL(2,c)$ consists of two-by-two unimodular matrices whose
elements are complex.  There are therefore six independent parameters,
and thus six generators of the Lie algebra.  This group is locally
isomorphic to the six-parameter Lorentz group or $O(3,1)$ applicable
to the Minkowskian space of three space-like directions and one
time-like direction.

Like the Lorentz group, the $SL(2,c)$ has a number of interesting
subgroups.  The subgroup most familiar to us is $SU(2)$ which is
locally isomorphic to the three-dimensional rotation group.  In
addition, this group contains three subgroups which are locally
isomorphic to the group $O(2,1)$ applicable to the Minkowskian
space of two space-like and one time-like dimensions.

One of the subgroups of $SL(2,c)$ is $SL(2,r)$ consisting of matrices
with real elements.  This subgroup is also called the $Sp(2)$ group
which we used in this paper in order to carry out the Iwasawa
decomposition.
Another interesting subgroup is the one we used for computing the $S$
matrix, which starts with the boundary matrix of Eq.(\ref{bd11})
and the phase-shift matrix of Eq.(\ref{ps11}).  This group is called
$SU(1,1)$.  The present paper exploits the isomorphism between
$Sp(2)$ and $SU(1,1)$.  While the physical world is describable in
terms of $SU(1,1)$, we carry out the Iwasawa decomposition in the
$Sp(2)$ regime.

Indeed, the conjugate transformation of Eq.(\ref{su11}) to
Eq.(\ref{sp2}) is from $SU(1,1)$ to $Sp(2)$, while the transition
from Eq.(\ref{sp2}) to Eq.(\ref{abcd}) is within the $Sp(2)$ group.
Thus, the transition from Eq.(\ref{su11}) to Eq.(\ref{abcd}) is a
conjugate transformation from the $SU(1,1)$ subgroup to the subgroup
$Sp(2)$ of $SL(2,r)$.

In this paper, we are concerned with the
decomposition of the $Sp(2)$ and $SU(1,1)$ matrices.
Unlike the traditional approach to group theory which starts from
the generators of the Lie algebra, we used in this paper an approach
similar to what Goldstein did for the three-dimensional rotation
group in terms of the Euler angles~\cite{gold80}.  There are
three-generators for the rotation group, but Goldstein starts with
rotations around the $z$ and $x$ directions.  Rotations around the
$y$ axis and the most general form for the rotation matrix can be
constructed from repeated applications of those two starting matrices.
Let us call this type of approach the ``Euler construction.''

There are three basic advantages of this approach.  First, the number
of ``starter'' matrices is less than the number of generators.
For example, we need only two starters for the three-parameter
rotation group.  In our case, we started with two matrices for the
three-parameter group $Sp(2)$ and also for $SU(1,1)$.
Second, each starter matrix takes a simple form and has its own
physical interpretation.

The third advantage can be stated in the following way.  Repeated
applications of the starter matrices will lead to a very complicated
expression.  However, the complicated expression can decomposed
into the minimum number of starter matrices.  For example, this
number is three for the three-dimensional rotation group.  This
number is also three for $SU(2)$ and $Sp(2)$.  We call this the
Euler decomposition.  The present paper is based on both the Euler
construction and the Euler decomposition.

Among the several useful Euler decompositions, the Iwasawa
decomposition plays an important role in the Lorentz group.   We have
seen in this paper what the decomposition does to the
two-by-two matrices of $Sp(2)$, but it has been an interesting
subject since Iwasawa's first publication on this
subject~\cite{iwa49}.  It is beyond the scope of this paper to
present a historical review of the subject.  However, we would
like to point out that there are areas of physics where this
important mathematical theorem was totally overlooked.

For instance, in particle theory, Wigner's little groups dictate the
internal space-time symmetries of massive and massless particles which
are locally isomorphic to $O(3)$ and $E(2)$ respectively~\cite{wig39}.
The little group is the maximal subgroup of the Lorentz group whose
transformations do not change the four-momentum of a given
particle~\cite{hkn00}.  The $E(2)$-like subgroup for massless
particles is locally isomorphic to the subgroup of $SL(2,c)$ which
can be started from one of the matrices in Eq.(\ref{shear1}) and the
diagonal matrix of Eq.(\ref{ps11}).  Thus there was an underlying
Iwasawa decomposition while the the $E(2)$-like subgroup was
decomposed into rotation and boost matrices~\cite{hks87jm}, but the
authors did not know this.  One of those authors is one of the
authors of the present paper.

In optics, there are two-by-two matrices with one vanishing
off-diagonal element.  It was generally known that this has something
to do with the Iwasawa effect, but Simon and Mukunda~\cite{simon98}
and Han {\it et al.}~\cite{hkn99} started treating the Iwasawa
decomposition as the main issue in their papers on polarized light.

In para-axial lens optics, the matrices of the form given in
Eq.(\ref{shear1}) are the starters~\cite{sudar85}, and repeated
applications of those two starters will lead to the most general
form of $Sp(2)$ matrices.  It had been a challenging problem since
1985~\cite{sudar85} to write the most general two-by-two matrix in
lens optics in terms the minimum number of those starter matrices.
This problem has been solved recently~\cite{bk01}, and the
central issue in the problem was the Iwasawa decomposition.

\end{document}